\documentclass[aps,prb,amsmath,amsfonts,amssymb,twocolumn,showpacs]{revtex4}
\usepackage{graphicx,psfrag,color}
\usepackage{dcolumn}
\usepackage{epsfig}

\newcommand{\im}{\mathrm i}

\newcommand{\llg}{\operatorname{ln}}

\newcommand{\eq}{\begin{equation}}
\newcommand{\en}{\end{equation}}
\newcommand{\bear}{\begin{eqnarray}}
\newcommand{\ear}{\end{eqnarray}}
\newcommand{\bt} { \begin{tabular} }
\newcommand{\et}{ \end{tabular} }
\newcommand{\bc} { \begin{center} }
\newcommand{\ec}{ \end{center} }

\begin{document}

\title{Magnetocaloric effect in integrable spin-$s$ chains}

\author{G.A.P. Ribeiro}
\email{pavan@df.ufscar.br}
\affiliation{Departamento de F\'{i}sica, Universidade Federal de S\~ao Carlos, S\~ao Carlos-SP 13565-905,
Brazil}

\date{\today}

\begin{abstract}
We study the magnetocaloric effect for the integrable antiferromagnetic high-spin chain. We present an exact computation of the Gr\"uneisen parameter, which is closely related to the magnetocaloric effect, for the quantum spin-$s$ chain on the thermodynamical limit by means of Bethe ansatz techniques and the quantum transfer matrix approach.  We have also calculated the entropy $S$ and the isentropes in the $(H,T)$ plane. We have been able to identify the quantum critical points  $H_c^{(s)}=\frac{2}{s+\frac{1}{2}}$ looking at the isentropes and/or the characteristic behaviour of the Gr\"uneisen parameter.
\end{abstract}

\maketitle

\section{Introduction}

The magnetocaloric effect has been known for more than hundred years \cite{WARBURG} 
and it is related to the temperature change of magnetic systems induced by an adiabatic 
variation of the external magnetic field. In recent years it has received considerable 
attention in view of new potential cooling applications \cite{TISHIN}. 

Many different 
families of magnetic materials ranging from ferromagnetic, ferrimagnetic to antiferromagnetic systems 
have been shown to present large or unusual magnetocaloric effect \cite{TSOKOL}. 
In particular, the magnetocaloric effect
has been measured in (quasi) one-dimensional materials which behave as quantum spin-$1/2$ 
chains \cite{TSUI} and high-spin chains \cite{SOSIN}. 

Moreover the existence of the magnetocaloric 
effect in one-dimensional systems has been studied theoretically
for spin-$1/2$ Heisenberg chain \cite{HONECKER1}, for spin-$s$ chain \cite{HONECKER2} and for mixed-spin chains\cite{BOSTREM} by means of 
numerical calculations. There are also some exact results for the $XY$ chain in transverse field\cite{HONECKER1} and 
for the Ising model \cite{HONECKER1,ROSCH}. Recently the magnetocaloric effect and the isentropes in the magnetic field/temperature $(H,T)$ plane
have been obtained exactly by Bethe ansatz techniques for the integrable spin-$1/2$ Heisenberg chain \cite{TRIPPE}.

The magnetocaloric effect $\left(\frac{\partial T}{\partial H}\right)_S=-\frac{\left(\partial S/\partial H\right)_T}{\left(\partial S/\partial T\right)_H}$ and the related quantity called Gr\"uneisen parameter $\Gamma_H$ have been pointed out as an important tool to detect and classify quantum critical points\cite{ROSCH0,ROSCH}. The Gr\"uneisen parameter for a magnetic systems can be written 
\eq
\Gamma_H=\frac{1}{T}\left(\frac{\partial T}{\partial H}\right)_S=-\frac{1}{C_H}\left(\frac{\partial M}{\partial T} \right)_H,
\en
where $C_H$ is the specific heat at a constant magnetic field and $\left(\frac{\partial M}{\partial T} \right)_H$ is the temperature variation of the magnetization $M$. This parameter $\Gamma_H$ has a characteristic sign change close to the quantum critical point, which is due to the accumulation of entropy at the critical point\cite{ROSCH}.

The integrable spin-$s$ generalization of the Heisenberg model\cite{KULISH} was exactly solved long ago \cite{BABUJIAN} providing all the eigenvalues and eigenvectors in terms of the Bethe equations. Its thermodynamic properties have been firstly studied\cite{BABUJIAN,SACRAMENTO} by means of the thermodynamic Bethe ansatz (TBA) method\cite{TBA}, which consist of an infinite number of non-linear integral equations (NLIE) for the free-energy. Alternatively, using the quantum transfer matrix (QTM) approach\cite{QTM}, it was derived a finite number of NLIE\cite{SUZUKI} which is more suitable for practical calculations.

Here we are interested in the exact computation of the magnetocaloric effect ($\Gamma_H$), entropy and the isentropes in the $(H,T)$ plane for
the integrable antiferromagnetic spin-$s$ chain. The thermodynamic quantities required to achieve this goal, like entropy, specific heat and magnetization, are determined as a function of temperature and magnetic field by means of the solution of a finite set of NLIE which arises from the QTM approach\cite{SUZUKI}.
 
This paper is organized as follows. In section \ref{INTEGRA}, we outline the integrable Hamiltonians and the associated integral equations. In section \ref{MAGNETO}, we present our results for the magnetocaloric effect and the isentropes in the $(H,T)$ plane. Our conclusions are given in
section \ref{CONCLUSION}.

\section{Hamiltonian and integral equations}\label{INTEGRA}

The Hamiltonian of the integrable spin-$s$ generalization of the Heisenberg model for $s=1/2,1$ and $3/2$ are given by
\bear
{\cal H}^{(\frac{1}{2})}&=&J\sum_{i=1}^{L}[\frac{1}{4}+\vec{S}_i \cdot \vec{S}_{i+1}], \\
{\cal H}^{(1)}&=&\frac{J}{4}\sum_{i=1}^{L} [3+ \vec{S}_i \cdot \vec{S}_{i+1} - (\vec{S}_i \cdot \vec{S}_{i+1})^2], \\
{\cal H}^{(\frac{3}{2})}&=&\frac{J}{532}\sum_{i=1}^{L} [234-27\vec{S}_i \cdot \vec{S}_{i+1} \nonumber \\
 &+& 8(\vec{S}_i \cdot \vec{S}_{i+1})^2+ 16(\vec{S}_i \cdot \vec{S}_{i+1})^3],	 
\ear
where $L$ is the number of sites and $\vec{S}_{i}=(\hat{S}_{i}^{x},\hat{S}_{i}^{y},\hat{S}_{i}^{z})$ are the $SU(2)$ generators.

One can write a closed formula for the Hamiltonian assuming a generic spin-$s$ value a follow,
\eq
{\cal H}^{(s)}=\frac{J}{2}\sum_{i=1}^{L} Q_s(\vec{S}_i \cdot \vec{S}_{i+1}) - H \sum_{i=1}^L \hat{S}_i^z,
\label{Hamiltonian}
\en
where
\eq
Q_s(x)=\sum_{j=0}^{2s}\left[2\psi(j+1)-\psi(1) -\psi(2s+1)\right] \prod_{\stackrel{k=0}{k\neq j}}^{2s} \frac{x-x_k}{x_j-x_k},
\en
with $x_{k}=\frac{1}{2}\left[k(k+1)-2s(s+1)\right]$, $\psi(x)$ the digamma function and $J$ is the exchange constant. From now on we assume $J=1$.  Note that we have also added a Zeeman term on the Hamiltonian (\ref{Hamiltonian}).

The free-energy of the system per lattice site calculated at the thermodynamic limit ($L\rightarrow \infty$) is given by
\eq
f(T,H)=f_0 -T \left( K\ast \llg{B\bar{B}} \right)(0),
\label{free-energy}
\en
where $f_0=\psi(2s+1)-\psi(\frac{2s+1}{2})+\psi(\frac{1}{2})-\psi(1)$, $K(x)=\frac{\pi}{\cosh{\left[ \pi x \right]}}$  and the symbol $\ast$
denotes convolution $f*g(x)=\int_{-\infty}^{\infty} f(x-y)g(y)dy$. 

\begin{widetext}

The auxiliary functions $b(x)$, $\bar{b}(x)$ and its simply related functions $B(x)=b(x)+1$ and $\bar{B}(x)=\bar{b}(x)+1$ are solution of the following set of non-linear integral equations\cite{SUZUKI}
\eq
\left(
\begin{array}{c}
\llg{y^{(\frac{1}{2})}(x)} \\
\vdots \\
\llg{y^{(s-\frac{1}{2})}(x)} \\
\llg{b(x)} \\
\llg{\bar{b}(x)}
\end{array}\right)=
\left(\begin{array}{c}
0 \\
\vdots \\
0 \\
- \beta d(x) + \beta \frac{H}{2}  \\
- \beta d(x) - \beta \frac{H}{2}
\end{array}\right)
+
{\cal K}*
\left(\begin{array}{c}
\llg{Y^{(\frac{1}{2})}(x)} \\
\vdots \\
\llg{Y^{(s-\frac{1}{2})}(x)} \\
\llg{B(x)} \\
\llg{\bar{B}(x)}
\end{array}\right),
\label{NLIE}
\en
where $d(x)=\frac{\pi}{2\cosh{\left[ \pi x \right]}}$, $\beta=1/T$ is the inverse of temperature and $H$ is the magnetic field.

The kernel matrix is given explicitly by
\eq
{\cal K}(x)= 
\left(
\begin{array}{cccccccc}
0 & K(x) & 0 & \cdots & 0 & 0 & 0 & 0 \\
K(x) & 0 &  K(x) &   & \vdots & \vdots & \vdots & \vdots  \\
 0 & K(x) & 0  &  &  & 0 & 0 & 0 \\
\vdots &  &   &   & 0 & K(x) & 0 & 0 \\
0 & 0 & \cdots  & 0 & K(x) & 0 & K(x) & K(x) \\
0 & 0 & \cdots  & 0 & 0 & K(x) & F(x) & -F(x+\im) \\
0 & 0 & \cdots  & 0 & 0 & K(x) & -F(x-\im) & F(x) \\
\end{array}\right),
\label{Kernel-x}
\en
which is a matrix of dimension $(2s+1)\times (2s+1)$ with $F(x)=\int_{-\infty}^{\infty}\frac{e^{-|k|/2+\im k x}}{2 \cosh{\left[ k/2 \right]}} dk $.

In order to obtain the desired thermodynamical quantities, we can calculate the derivatives of the free-energy with respect to temperature $T$ (or more conveniently $\beta$) and magnetic field $H$. It turns out to be more efficient to calculate the derivatives of the free-energy in terms of the solution of linear integral equations. These equation are obtained by differentiation of the equation (\ref{NLIE}). This way we can avoid numerical differentiation of the free-energy.

Specifically one can write the entropy $S=-\left(\frac{\partial f}{\partial T}\right)_H=\beta^2\left(\frac{\partial f}{\partial \beta}\right)_H$ as follow,
\eq
S=(K\ast \ln{B \bar{B}})(0) - \beta (K\ast \partial_{\beta} \ln{B\bar{B}})(0),
\label{entropy}
\en
where $\partial_{\beta}\ln{B(x)}=\frac{b(x)}{b(x)+1}[\partial_{\beta}\ln{b(x)}]$,   $\partial_{\beta}\ln{\bar{B}(x)}=\frac{\bar{b}(x)}{\bar{b}(x)+1}[\partial_{\beta}\ln{\bar{b}(x)}]$ and  $\partial_{\beta}\ln{Y^{(j)}(x)}=\frac{y^{(j)}(x)}{y^{(j)}(x)+1}[\partial_{\beta}\ln{y^{(j)}(x)}]$ for  $ j=1/2,\dots,s-1/2$ . These new auxiliary functions are solution of the following system of linear integral equations
\eq
\left(
\begin{array}{c}
\partial_{\beta}\llg{y^{(\frac{1}{2})}(x)} \\
\vdots \\
\partial_{\beta}\llg{y^{(s-\frac{1}{2})}(x)} \\
\partial_{\beta}\llg{b(x)} \\
\partial_{\beta}\llg{\bar{b}(x)}
\end{array}\right)=
\left(\begin{array}{c}
0 \\
\vdots \\
0 \\
-  d(x) + \frac{H}{2}  \\
-  d(x) - \frac{H}{2}
\end{array}\right)
+
{\cal K}*
\left(\begin{array}{c}
\partial_{\beta}\llg{Y^{(\frac{1}{2})}(x)} \\
\vdots \\
\partial_{\beta}\llg{Y^{(s-\frac{1}{2})}(x)} \\
\partial_{\beta}\llg{B(x)} \\
\partial_{\beta}\llg{\bar{B}(x)}
\end{array}\right).
\label{LIE1}
\en
To obtain the entropy in the $(H,T)$ plane, one has to solve the above equations (\ref{NLIE}) and (\ref{LIE1}) varying the temperature and the magnetic field.

The specific heat $C_H=T\left(\frac{\partial S}{\partial T}\right)_H=-\beta \left(\frac{\partial S}{\partial \beta}\right)_H$ can be obtained from  (\ref{entropy}), 
\eq
C_H= \beta^2 (K\ast \partial_{\beta}^2 \ln{B\bar{B}})(0),
\label{specificheat}
\en
which is given in terms of the solution of following linear integral equations
\eq
\left(
\begin{array}{c}
\partial_{\beta}^2\llg{y^{(\frac{1}{2})}(x)} \\
\vdots \\
\partial_{\beta}^2\llg{y^{(s-\frac{1}{2})}(x)} \\
\partial_{\beta}^2\llg{b(x)} \\
\partial_{\beta}^2\llg{\bar{b}(x)}
\end{array}\right)=
{\cal K}*
\left(\begin{array}{c}
\partial_{\beta}^2\llg{Y^{(\frac{1}{2})}(x)} \\
\vdots \\
\partial_{\beta}^2\llg{Y^{(s-\frac{1}{2})}(x)} \\
\partial_{\beta}^2\llg{B(x)} \\
\partial_{\beta}^2\llg{\bar{B}(x)}
\end{array}\right),
\label{LIE2}
\en
where $\partial_{\beta}^2\llg{B(x)}=\frac{b(x)}{b(x)+1} \left\{\frac{[\partial_{\beta}\llg{b(x)}]^2}{b(x)+1} + [\partial_{\beta}^2\llg{b(x)}] \right\}$  and  $\partial_{\beta}^2\llg{Y^{(j)}(x)}=\frac{y^{(j)}(x)}{y^{(j)}(x)+1} \left\{\frac{[\partial_{\beta}\llg{y^{(j)}(x)}]^2}{y^{(j)}(x)+1} + [\partial_{\beta}^2\llg{y^{(j)}(x)}] \right\}$.

\end{widetext}   

In order to obtain the Gr\"uneisen parameter we have also to determine the $\left(\frac{\partial M}{\partial T}\right)_H$. Therefore we have firstly to calculate the magnetization $M=-\left(\frac{\partial f}{\partial H} \right)_T$ from (\ref{free-energy}),
\eq
M=\frac{1}{\beta} (K\ast \partial_{H} \ln{B\bar{B}})(0),
\label{magnetization}
\en
which is written in terms of the auxiliary function $\partial_{H}\ln{B(x)}=\frac{b(x)}{b(x)+1}[\partial_{H}\ln{b(x)}]$ (likewise for the other auxiliary functions) that are now derivatives with respect with the magnetic field.  These new auxiliary functions are solution of the following system of linear integral equations
\eq
\left(
\begin{array}{c}
\partial_{H}\llg{y^{(\frac{1}{2})}(x)} \\
\vdots \\
\partial_{H}\llg{y^{(s-\frac{1}{2})}(x)} \\
\partial_{H}\llg{b(x)} \\
\partial_{H}\llg{\bar{b}(x)}
\end{array}\right)=
\left(\begin{array}{c}
0 \\
\vdots \\
0 \\
 \frac{\beta}{2}  \\
-\frac{\beta}{2}
\end{array}\right)
+
{\cal K}*
\left(\begin{array}{c}
\partial_{H}\llg{Y^{(\frac{1}{2})}(x)} \\
\vdots \\
\partial_{H}\llg{Y^{(s-\frac{1}{2})}(x)} \\
\partial_{H}\llg{B(x)} \\
\partial_{H}\llg{\bar{B}(x)}
\end{array}\right).
\en

The derivative of the magnetization with respect to temperature for constant magnetic field $\left(\frac{\partial M}{\partial T}\right)_H=-\beta^2 \left(\frac{\partial M}{\partial \beta}\right)_H$ can be finally obtained from (\ref{magnetization}), which results 
\eq
\frac{\partial M}{\partial T}= (K\ast \partial_{H} \ln{B\bar{B}})(0) -\beta (K\ast \partial_{\beta H}^2 \ln{B\bar{B}})(0),
\en
where $\partial_{\beta H}^2\ln{B(x)}=\frac{b(x)}{b(x)+1}\left\{\frac{ [\partial_{\beta}\ln{b(x)}] [\partial_{H}\ln{b(x)}]}{b(x)+1} + [\partial_{\beta H}^2\ln{b(x)}]  \right\}$ which should satisfies
\eq
\left(
\begin{array}{c}
\partial_{\beta H}^2\llg{y^{(\frac{1}{2})}(x)} \\
\vdots \\
\partial_{\beta H}^2\llg{y^{(s-\frac{1}{2})}(x)} \\
\partial_{\beta H}^2\llg{b(x)} \\
\partial_{\beta H}^2\llg{\bar{b}(x)}
\end{array}\right)=
\left(\begin{array}{c}
0 \\
\vdots \\
0 \\
\frac{1}{2}  \\
-\frac{1}{2}
\end{array}\right)
+
{\cal K}*
\left(\begin{array}{c}
\partial_{\beta H}^2\llg{Y^{(\frac{1}{2})}(x)} \\
\vdots \\
\partial_{\beta H}^2\llg{Y^{(s-\frac{1}{2})}(x)} \\
\partial_{\beta H}^2\llg{B(x)} \\
\partial_{\beta H}^2\llg{\bar{B}(x)}
\end{array}\right).
\en

\section{Gr\"uneisen parameter and entropy}
\label{MAGNETO}

In this section we will present the results for the Gr\"uneisen parameter, which is closely related to the magnetocaloric effect. We will also show the results for the entropy and the isentropes in the $(H,T)$ plane.

\begin{figure}[tb!]
\includegraphics[width=\columnwidth]{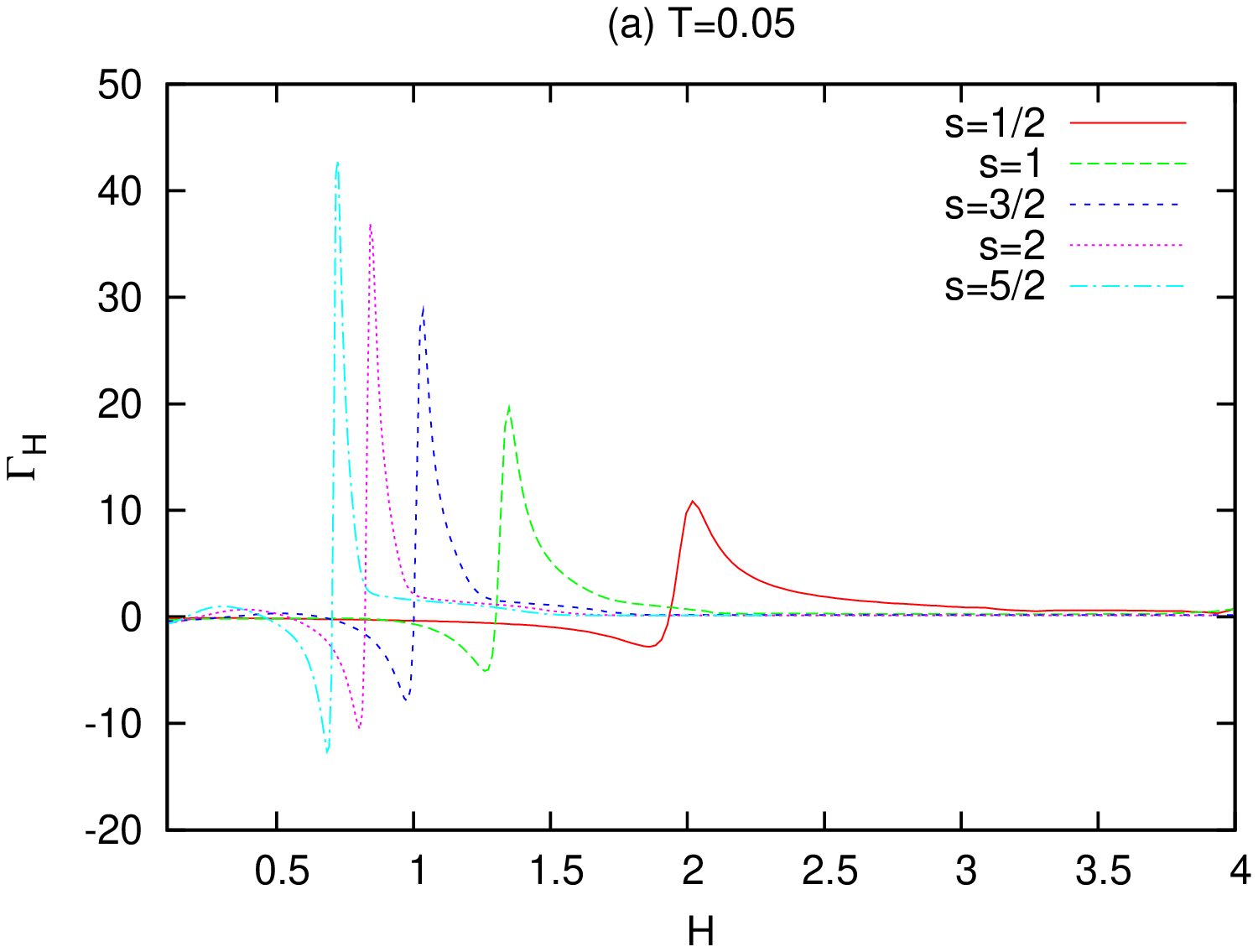}
\includegraphics[width=\columnwidth]{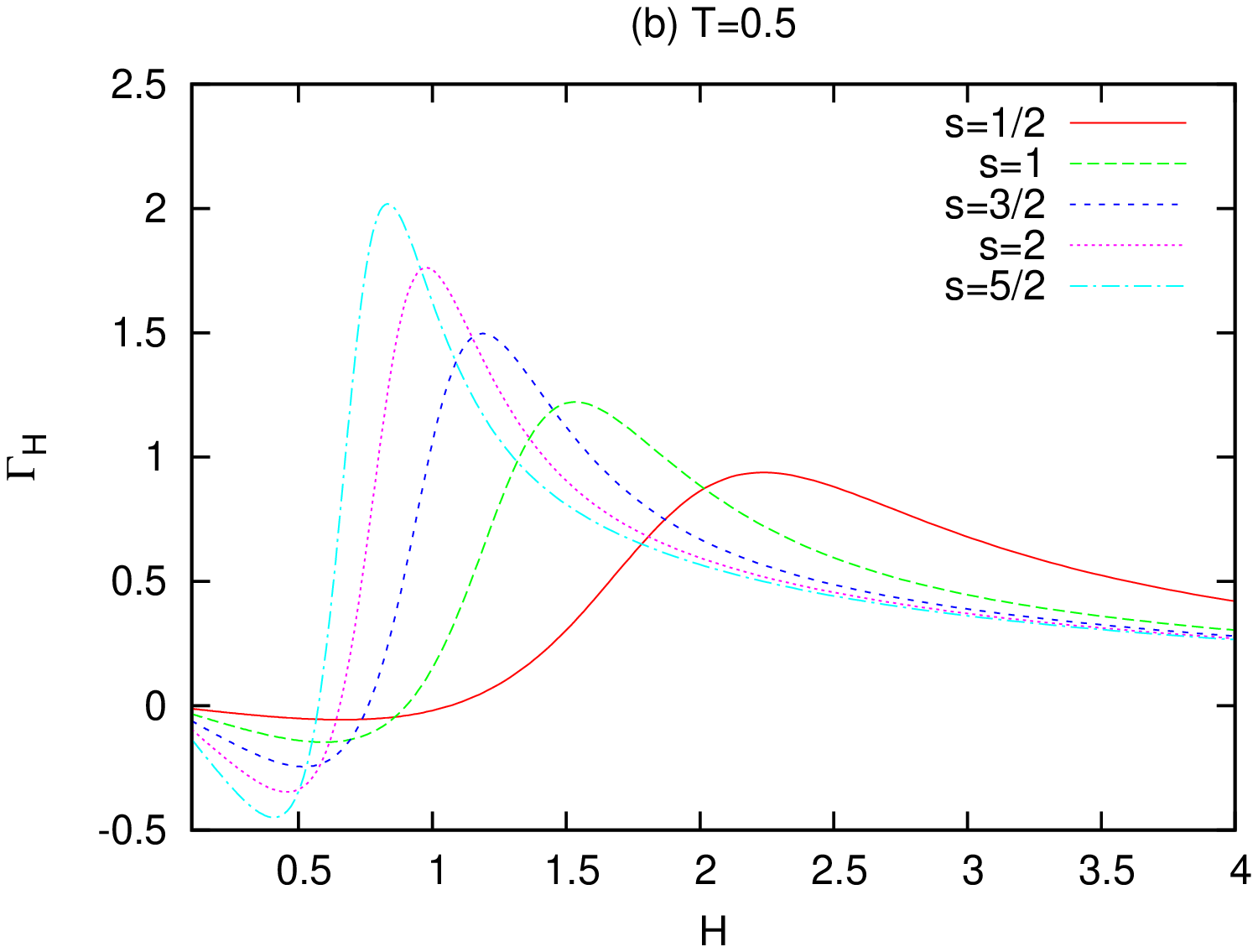}
\includegraphics[width=\columnwidth]{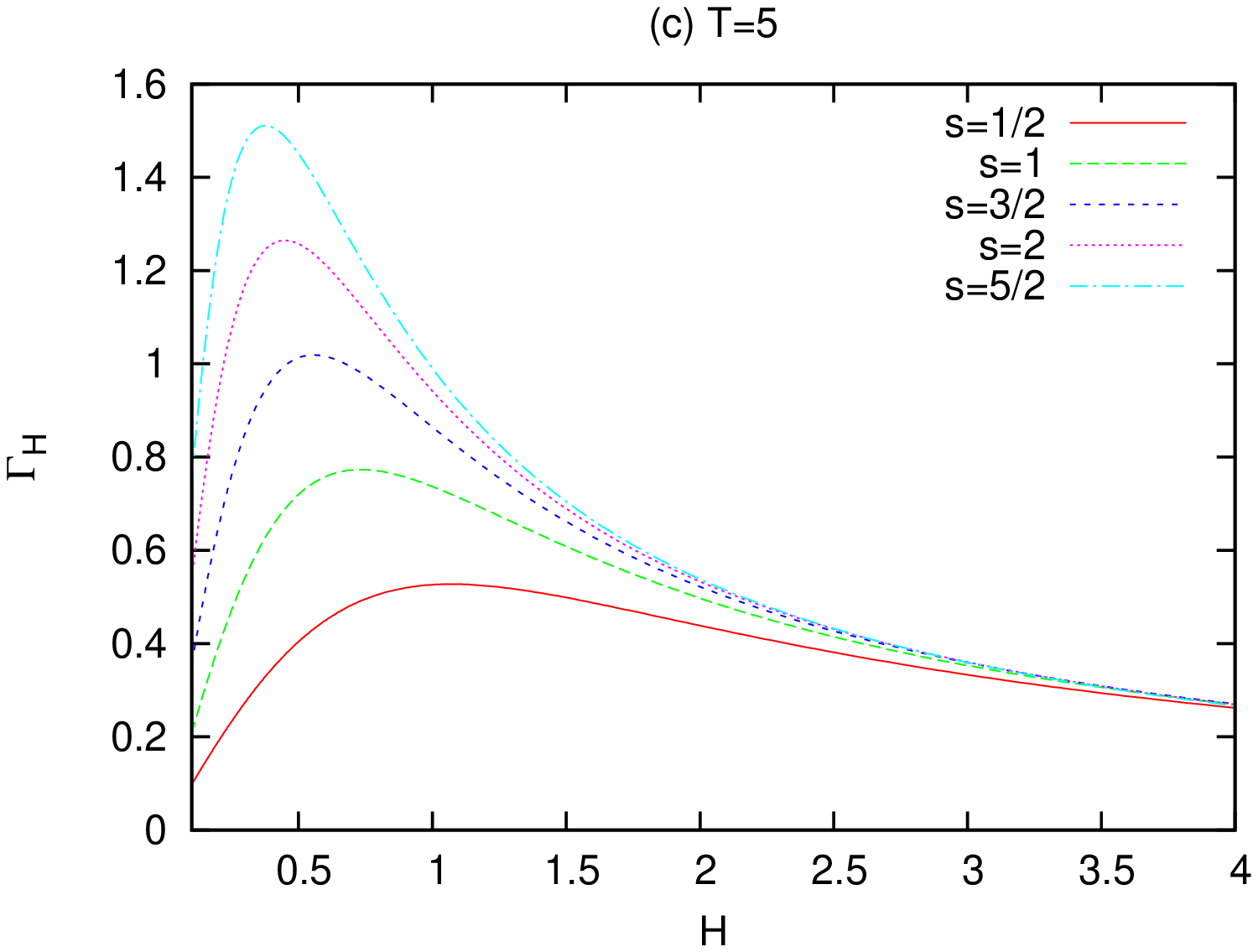}
\caption{(Color online) The Gr\"uneisen parameter $\Gamma_H$ for different values of the spin for (a) $T=0.05$, (b) $T=0.5$ and (c) $T=5$.
\label{figure1}}
\end{figure}

First we show on Fig. (\ref{figure1}) the Gr\"uneisen parameter $\Gamma_H$ as a function of magnetic field for some spin values $s=1/2,1,3/2,2$ and $5/2$. Note that the case $s=1/2$ was calculated on Ref. 
10. For the temperature $T=0.05$, we can see that the transition to saturation at $H^{(s)}_c=\frac{2}{s+\frac{1}{2}}$ is signaled by sign changes of the Gr\"uneisen parameter from negative to positive values toward the higher fields values. For higher temperatures, like $T=0.5$, we note that these sign changes move away from the zero temperature saturation field, which separates the antiferromagnetic and ferromagnetic phases. Finally, if we go further to higher temperatures, e.g  $T=5$ one can see that all the characteristic behaviour have disappeared, which imply that the thermal fluctuations are already strong enough to drive the system to excited states where no quantum phase transition effects can be seen. Moreover there is a small structure at low magnetic fields and low temperatures Fig. \ref{figure1}(a) which is due to the singular nature of the point $(H=0)$ of the isotropic integrable spin chains\cite{TRIPPE}.

\begin{figure}[tb!]
\includegraphics[width=\columnwidth]{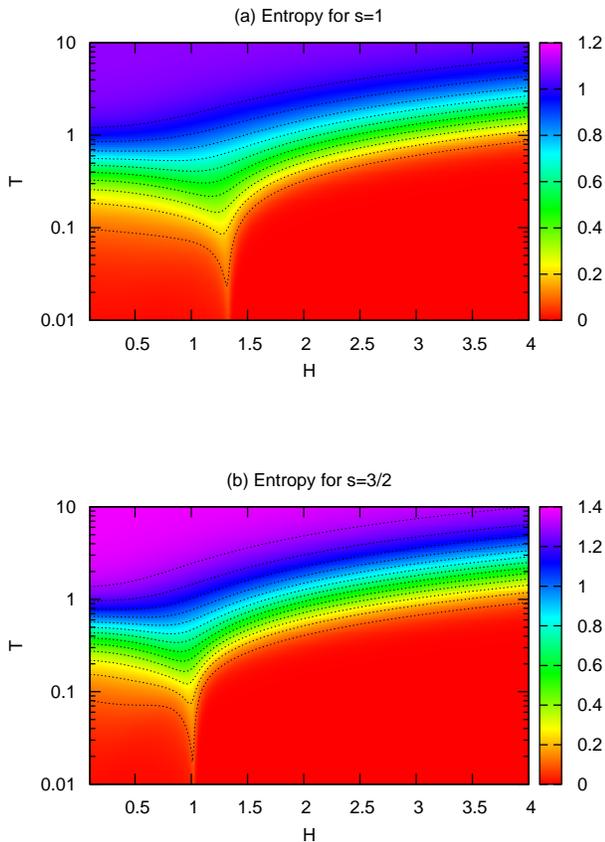}
\caption{(Color online) Entropy $S(H,T)$ for $s=1,3/2$. The isentropes are for $S=0.1, 0.2, \dots, 1.3$.
\label{figure2}}
\end{figure}

\begin{figure}[tb!]
\includegraphics[width=\columnwidth]{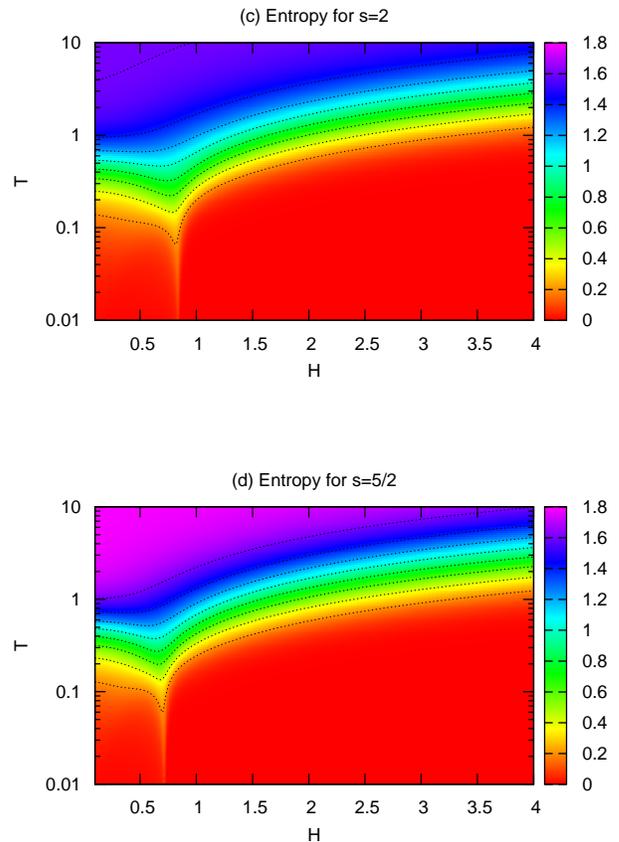}
\caption{(Color online) Entropy $S(H,T)$ for $s=2,5/2$. The isentropes are for $S=0.1, 0.2, \dots, 1.6$.
\label{figure3}}
\end{figure}

We show on Figs. \ref{figure2} and \ref{figure3} the entropy and the isentropes for the spin-$s$ chain in the $(H,T)$ plane for $s=1,3/2,2$ and $5/2$. 

The quantum phase transitions are indicated by the isentropes which are tilted towards the quantum critical point\cite{ROSCH} showing a minima nearby $H_c^{(s)}$ or equivalently the entropy peaks at the critical point. This accumulation of entropy nearby the critical point indicates that the systems is maximally undecided which ground state to choose\cite{ROSCH}. Moreover the Gr\"uneisen parameter, which is proportional to the slope of the isentropes $(\frac{\partial T}{\partial H})$, has a different sign on each side of the quantum critical point as we have shown on Fig. \ref{figure1}. Besides that the isentropes are very steep nearby the critical point indicating the existence of a large magnetocaloric effect.

\section{Conclusion}
\label{CONCLUSION}

In this paper we have studied the magnetocaloric effect for the integrable spin-$s$ chain. We have calculated the Gr\"uneisen parameter, which is proportional to the magnetocaloric effect, as a function of the external magnetic field  on the thermodynamic limit and at finite temperatures. We have also obtained entropy and the isentropes in the $(H,T)$ plane. 

The quantum critical point $H_c^{(s)}$ have been identified by the minima of the isentropes and by the sign changes of the Gr\"uneisen parameter as a function of the magnetic field. Our results are in agreement with the previous results for the $s=1/2$ case\cite{TRIPPE}.  

We hope that our exact results could be useful for understanding experimental results for quasi one-dimensional systems, e.g \cite{SOSIN}. We also expect that our results could be further extended to the case of alternating spin-$(S_1,S_2)$ chain \cite{RIBEIRO}.

\begin{acknowledgments}
The author thanks FAPESP for financial support.
\end{acknowledgments}

\end{document}